\documentclass[article]{revtex4}
\usepackage{graphicx}

\begin{document}

\title{Nonlinear surface magneto-plasmonics in Kretschmann multilayers}


\author{Ilya Razdolski\footnote{Present address: Fritz-Haber-Institut der MPG, Phys. Chemie, Faradayweg 4-6, 14195 Berlin, Germany}, Andrei Kirilyuk, Theo Rasing}
\affiliation{Institute for Molecules and Materials, Radboud University Nijmegen,
6525 AJ Nijmegen, Netherlands}

\author{Denys Makarov, Oliver G. Schmidt}
\affiliation{Institute for Integrative Nanosciences, IFW Dresden, 01069
Dresden, Germany}

\author{Vasily V. Temnov}
\affiliation{Institut des Mol\'ecules et Mat\'eriaux du Mans, UMR
CNRS 6283, Universit\'e du Maine, 72085 Le Mans cedex, France}

\date{\today}

\maketitle \textbf{The nonlinear magneto-plasmonics
\cite{RazdolskiPRB13,ZhengSciRep14,KrutyanskiyPRB15} aims to utilize plasmonic
excitations to control the mechanisms and taylor the efficiencies
of the non-linear light frequency conversion at the nanoscale. We
investigate the mechanisms of magnetic second harmonic generation
in hybrid gold-cobalt-silver multilayer structures, which support
propagating surface plasmon polaritons at both fundamental and
second harmonic frequencies. Using magneto-optical spectroscopy in
Kretschmann geometry, we show that the huge magneto-optical
modulation of the second harmonic intensity is dominated by the excitation of surface
plasmon polaritons at the second harmonic frequency, as shown by
tuning the optical wavelength over the spectral region of strong
plasmonic dispersion. Our proof-of-principle experiment highlights bright prospects of 
nonlinear magneto-plasmonics and contributes to the general understanding of
the nonlinear optics of magnetic surfaces and interfaces.}

Rapid development of plasmonics has facilitated an outstanding
progress in understanding, designing and controlling the optical
response of metallic nanostructures, including that in
nonlinear-optical domain
\cite{TemnovNatPhot10,KauranenNatPhot12,Armellesreview,BelotelovNatComm13}.
The optical second harmonic generation (SHG) from solid interfaces
represents a well-known experimental technique with
numerous applications in physics, chemistry and biology. Being a
natural tool to enhance the light-matter interaction,
the plasmon-assisted localization of the electric field in small
volumes proved to be very effective at elevating the efficiency of the
nonlinear-optical processes. Beyond the physics of plasmonic
nanoantennas requiring the fabrication of sophisticated metallic
nanoobjects \cite{NovotnyNatPhot11,HankeNano12}, the Kretschmann geometry for
the excitation of surface plasmon polaritons (SPPs) in a thin
metallic layer on a dielectric prism plays a special role due to
its robustness and simplicity\cite{Kretschmann}. A huge plasmonic enhancement of
the electric field caused by the phase-matched excitation of SPPs
under the resonant coupling conditions has been shown to boost the SHG
intensity
\cite{SimonPRL74,NaraokaOpCom05,GrossePRL12}.
The most recent experiments have evidenced that not only the SPP-resonance
at a fundamental frequency $\omega$ but also the excitation of a
second harmonic SPP at the doubled frequency $2\omega$ plays an
important role in the process of the nonlinear-optical conversion
\cite{PalombaPRL08,GrossePRL12}. However, due to their dispersive
nature a simultaneous excitation and coupling between the SPPs at
fundamental and second harmonic frequencies has not been reported
so far.

\begin{figure}[!ht]
  \includegraphics[width=0.7\textwidth]{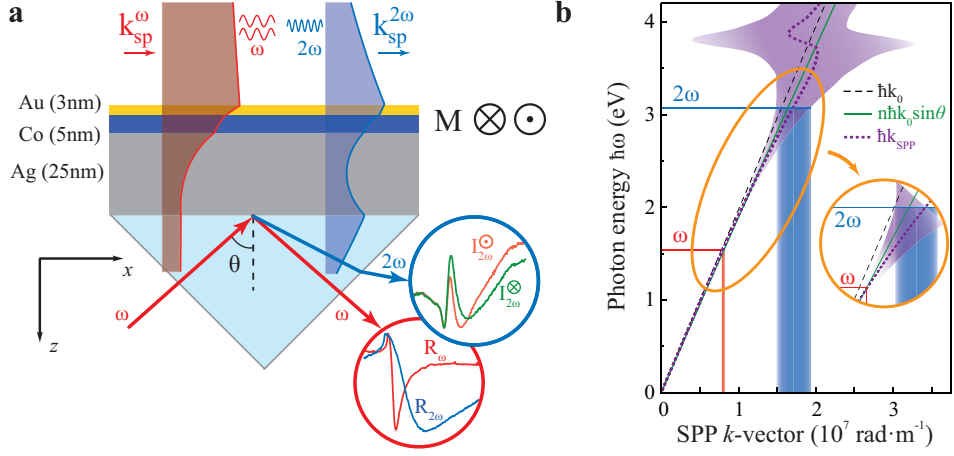} \caption{ 
	{\bf Surface plasmons in Au/Co/Ag multilayer structures.} 
{\bf a} Sketch of the experimental geometry showing the spatial distribution of the square of the normal projection of the electric field at the angles of incidence $\theta$ corresponding to the SPP excitations at the fundamental $\omega$ and double $2\omega$ frequencies. Insets in the coloured circles show the experimental angular dependencies of the linear-optical reflectivity and SHG intensities for the two opposite directions of magnetisation of the cobalt layer.
{\bf b} Dispersion of the SPP in the Au/Co/Ag trilayer under study. Black dash and green solid lines represent the photons in vacuum and in glass, respectively. Thick dot purple line is the calculated SPP dispersion, its linewidth is shown with the purple background area. The inset illustrates the possibility of a simultaneous excitation of the SPPs at both frequencies $\omega$ and $2\omega$.}
\end{figure}

The main idea of this study was to design a nanoscale optical
experiment where we could excite, detect and manipulate the
nonlinear interactions between the SPPs at the fundamental and SHG
frequencies. To achieve this goal we have adopted the concepts
of linear magneto-plasmonics in hybrid metal-ferromagnet
multilayer structures \cite{Armellesreview,TemnovNatPhot12} together with the
most recent ideas in nonlinear magneto-plasmonics
\cite{RazdolskiPRB13,KrutyanskiyPRB13,ChekhovOExp14,ZhengSciRep14,KrutyanskiyPRB15}. 
Our experimental geometry is sketched in Fig.~1a. A thin
gold/cobalt/silver trilayer structure was grown on a glass
substrate by means of the magnetron sputtering. A 5~nm-thin magneto-optically
active layer of ferromagnetic cobalt was protected from oxidation
by a 3~nm-thin layer of gold. A 25~nm-thick silver layer acted as
main constituent in this hybrid plasmonic nanostructure, which was
excited by collimated femtosecond laser pulses through the glass
prism (Kretschmann geometry). The reflected SHG intensity, as well as the linear reflectivity signal at both fundamental $\omega$ and double frequency $2\omega$ was recorded as a function of the incidence angle $\theta$ for the
two opposite in-plane directions of magnetization ($+M$ and $-M$ in
Fig.~1a) of cobalt in the transverse Kerr geometry.

Understanding the properties of the SPP dispersion \cite{DionnePRB05} in our
structure (see Fig.~1b and Supplementary Information for details)
is one of the key points for nonlinear magneto-plasmonics. Due to
inevitable dispersion, the SPP excitations at fundamental and
double frequencies in Kretschmann geometry occur at slightly
different angles, suggesting that simultaneous phase-matched
excitation of both SPPs is impossible. 
However, nonlinear-optical considerations account for the following nonlinear phase-matching condition
\begin{equation}\label{PhaseMatching}
2k_0(\omega) n(\omega)\sin\theta_{\rm nl}=k_{\rm spp}(2\omega)\,,
\end{equation}
between the excitation source at the silver-glass interface characterized
by the in-plane component of the $k$-vector $k^{\omega}_0\sin\theta$ and
the second harmonic SPP at the gold-air interface with the $k$-vector $k^{2\omega}_{\rm spp}$.
Being just one of several possible SPP frequency conversion pathways
\cite{GrossePRL12,HeckmannOpEx13}, this phase-matching condition plays a crucial
role in our interpretation as it determines the resonant SPP-induced
enhancement of the nonlinear susceptibility $\chi^{(2)}$.

The experimental results are summarized in Fig.~2. The linear optical reflectivity in Kretschmann configuration using the
fundamental (920~nm) and second harmonic wavelengths (460~nm,
frequency doubled in a BBO-crystal) are displayed in Fig. 2a. A
narrow plasmonic reflectivity dip at 920~nm observed at 42.5
degrees contrasts with a much broader minimum at 460~nm at $\theta\approx$ 47
degrees, in agreement with much higher SPP-losses at $2\omega$.
In terms of linear magneto-plasmonics these two resonances behave
similarly: the relative magnetisation-induced variations of the reflectivity,
$\Delta R/R$ neatly follow the angular differential reflectivity $dR/d\theta$,
suggesting that altering the magnetization direction induces a small shift of the reflectivity
spectrum due to the magnetic contribution to the SPP wavevector
\cite{GonzalezDiazPRB07,TemnovNatPhot10}.

\begin{figure}[!ht]
  \includegraphics[width=0.9\textwidth]{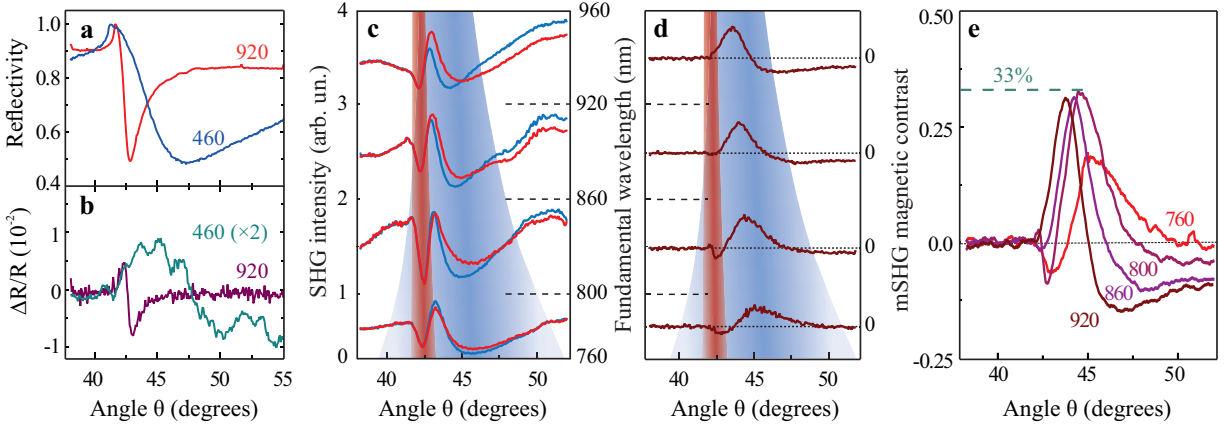} 
	\caption{
	{\bf Angular spectra of the linear and nonlinear magnetooptical response in presence of the SPPs.}
	{\bf a,b} Linear-optical reflectivity $R$ and magnetization-induced reflectivity variations $\Delta R/R$ for the excitation with the 920 and 460 nm wavelength. 
	{\bf c,d} SHG intensity and (d) mSHG contrast $\rho$ angular spectra plotted with an offset. Red and blue background areas illustrate the SPP dispersion in the experimental spectral range (760-920 nm for fundamental, 380-460 for the SHG, respectively.)
	{\bf e} Angular spectra of the mSHG magnetic contrast $\rho$ for the four fundamental wavelengths, no offset is introduced. Inset:  Illustration of the magnetisation-induced changes of the total SHG intensity, $I_{2\omega}(\pm M)\propto |\vec{E}^{2\omega}(\pm M)|^2
	=|\vec{E}^{2\omega}\pm \vec{E}^{2\omega}_{\rm m}|^2$.
	}
\end{figure}

Whereas the linear reflectivity at fundamental frequency $\omega$
shows only small modulations of the order of 1\%, in line with Ref.
\cite{Armellesreview}, the SHG angular spectra at $2\omega$ display
drastic changes, as shown in Fig.~2b for four different values of
the fundamental wavelength 760, 800, 820 and 920~nm. Red and blue
background areas show the SPP dispersion (see Fig.~1b) at the frequencies $\omega$ and $2\omega$, respectively,
recalculated as a function of incidence angle $\theta$. Over the SHG
wavelengths range (380-460~nm) the SPP shows a strongly dispersive behavior resulting in a shift of the resonant
angle and decrease of losses (see the linewidth). Fundamental SPP remains
nearly dispersionless and demonstrates no noticeable change of its
linewidth. Contrary to previously reported results on a gold film \cite{PalombaPRL08},
the angular positions of the SPP resonances for the fundamental and SHG
frequencies in our multilayer structure correspond to the pronounced minima in the SHG intensity.
A strong dependence of SHG intensity on magnetization
direction $I_{2\omega}(\pm M)$ is quantified by the magnetic SHG (mSHG) contrast $\rho$:
\begin{equation}\label{rho}
  \rho=\frac{I_{2\omega}(+M)-I_{2\omega}(-M)}{I_{2\omega}(+M)+I_{2\omega}(-M)},
\end{equation}
shown in Fig.~2d. Surprisingly, the largest magnetisation-induced modulation of the SHG intensity
is accompanied by the SPP excitation at the SHG frequency and not the fundamental one.
Angular dependence of mSHG contrast re-plotted in Fig.~2e without an offset displays the largest reported values of
modulation reaching 33~$\%$ around $\theta=44$~degrees. For the
shortest wavelength (760 nm) the mSHG maximum is only about 20$\%$, since in
this case the SPP damping at the SHG frequency becomes so large that the system approaches the region with the
non-propagating (propagation length $\lambda^{\rm SPP}<\lambda_{2\omega}$) SPPs. 
This observation links up our measurements to the most recent results by Zheng {\it et al.}
\cite{ZhengSciRep14}, who reported similar values of the mSHG
contrast on a 10~nm-thin iron film on glass, a structure not
supporting SPPs at the SHG frequency. This fact, along with the
dispersive shift of mSHG maximum reinforces our conclusion that
the 33~$\%$ large mSHG contrast (which is equivalent to the increase of
the SHG intensity by a factor of 2 upon magnetization reversal)
is dominated by the SPP resonance at the SHG frequency.

In order to understand these experimental observations we
have used the approach proposed by Palomba and Novotny \cite{PalombaPRL08},
who explained the complex angular dependence of the SHG intensity from
a thin gold film on glass by an interference of two contributions
coming from the gold-air and gold-glass interfaces. In our case (Fig.~3a)
the silver-sapphire interface acts as a source of the nonmagnetic
SHG $\vec{E}^{2\omega}_{1}$, and the upper part
consisting of Au and Co layers is assumed to generate the electric
field $\vec{E}^{2\omega}_{2}$ containing both magnetic $\chi^{(2m)}$ and
non-magnetic $\chi^{(2)}$ contributions. Thus the total SHG intensity $I_{2\omega}$ is described by:
\begin{equation}\label{intensity}
 I_{2\omega}\propto |\vec{E}^{2\omega}_1+\vec{E}^{2\omega}_2\pm \vec{E}^{2\omega}_{2m}|=
|\chi^{(2)}_1:\vec{E}_1 \vec{E}_1 +
\chi^{(2)}_2:\vec{E}_{2} \vec{E}_{2}\pm 
\chi^{(2m)}_2:\vec{E}_{2} \vec{E}_{2}|^2\,,
\end{equation}
where complex tensor components $\chi^{(2)}_1$ and
$\chi^{(2)}_2\pm \chi^{(2m)}_2$ represent
the effective optical nonlinearities at both interfaces.
The angular dependences of the electric fields $\vec{E}_1$ and
$\vec{E}_2$ at the fundamental frequency $\omega$, which are driving
the SHG process at both interfaces, are calculated
by finite difference time domain (FDTD) method for the 860~nm
excitation wavelength (see Fig.~3c-d). 

\begin{figure}[!h]
  \includegraphics[width=0.9\textwidth]{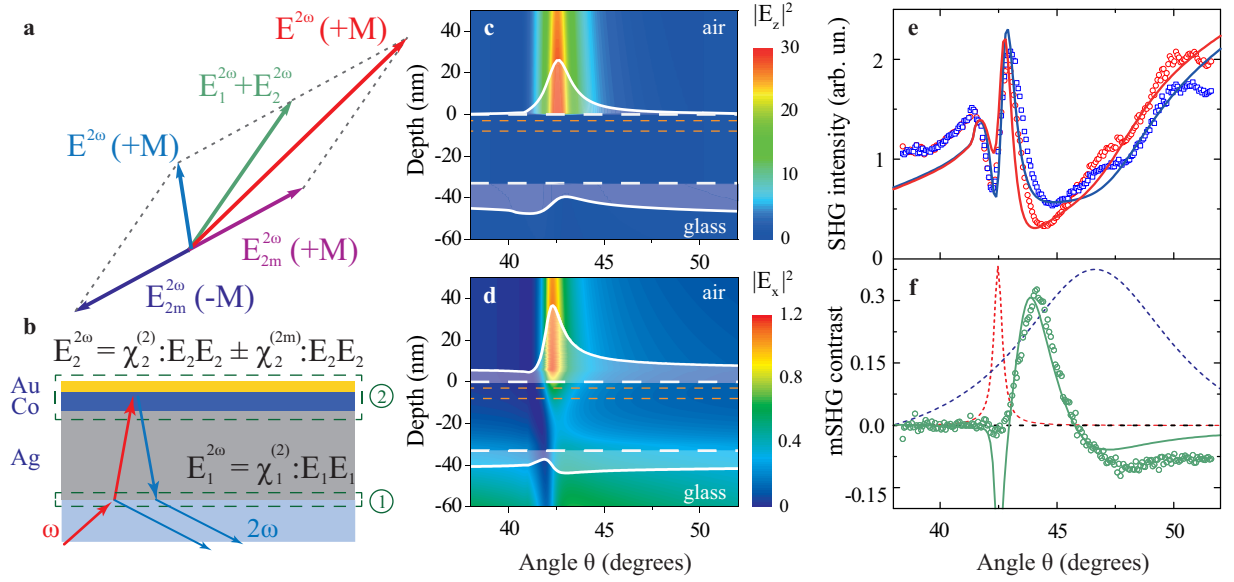}
	\caption{
	{\bf Schematics of the SPP-assisted mSHG generation in a multilayer structure.}
	{\bf a} Illustration of the origin of the inequality of the SHG intensities for the opposite direction of the external magnetic field $H$ leading to the non-zero magnetic contrast $\rho$.
	{\bf b} SHG sources at the two interfaces, bottom (1, glass/Ag) and top (2, air/Au/Co/Si).
	{\bf c-d} False colour spatial distributions of the square of the electric field projections $|E_z|^2$ and
	$|E_x|^2$ at the 860 nm fundamental wavelength. Dashed white lines show the sample borders. Solid white lines represent the corresponding $|E|^2$ angular distributions at the top and bottom interfaces. 
	{\bf e-f} SHG intensity and mSHG contrast angular spectra (open circles and squares) for the 860 nm fundamental wavelength together with the fit lines based on the Eq.~(\ref{intensity}). 
Dashed lines are the angular profiles corresponding to the resonant SPP excitation at the frequency $\omega$ and nonlinear SPP excitation $\omega \rightarrow 2\omega$ (see Fig.~5S in the Supplementary Information).
	}
\end{figure}

Based on these calculations, the modulation of SHG intensity in the vicinity of
fundamental SPP resonance can be explained by destructive
interference of two contributions $\vec{E}^{2\omega}_{1}$ and
$\vec{E}^{2\omega}_{2}$ \cite{PalombaPRL08}. For our multilayer
structure the amplitude of the SHG light generated at the upper
interface and transmitted through the prism at any angle $\theta$
turns out to be smaller as compared to the other one:
$|\vec{E}^{2\omega}_{2}|<|\vec{E}^{2\omega}_{1}|$. This is why the
strong increase of $\vec{E}_2$ in Fig.~3c-d at the top interface explains the SHG intensity
minimum around the fundamental SPP resonance at $\theta=42.5~$degrees.

The same explanation holds for the SHG minimum around
$\theta=45~$degrees albeit for a different reason. 
We note that the nonlinear susceptibility $\chi^{(2)}$ is known to acquire a SPP-induced resonant contribution, $\chi^{(2)}=\chi^{(2)}_{\rm nr}+\chi^{(2)}_{\rm res}$ \cite{RaschkeCPL02,Heinz} with:
\begin{equation}\label{chires}
\chi^{(2)}_{\rm res}(\theta)\propto\frac{1}{\theta-\theta_{\rm nl}+i\Gamma}\,.
\end{equation}
Owing to this resonant contribution, the SHG field enhanced at the top interface destructively interferes with the one generated at the bottom interfaces, which explains the experimentally observed SHG intensity minimum.
In order to understand the angular spectrum of this resonant contribution we approximated it with the Lorentzian resonance line. The width $\Gamma$ and the angular position $\theta_{\rm nl}$ of this resonance line was chosen according to the calculations shown in Fig.~1b for the nonlinear SPP excitation at frequency $2\omega$, or 430~nm wavelength. This resonant shape successfully accounts for the large mSHG modulation at the angles corresponding to the nonlinear SPP excitation quantified by Eq.~(1). Based on Eq.~(3) we were able to fit the experimental angular spectra using the resonant part of the nonlinear susceptibility $\chi^{(2)}_{\rm res}$ only.

Leaving out the details of our fitting of Eq.~(3) for the Supplementary Information, where the choice of predominant $\chi^{(2)}$ components (from the six non-zero ones, three magnetic and three non-magnetic, at each interface) also justified,
we would like to claim to have obtained a good quantitative
agreement between the theory and experiment both for angular
dependence of the SHG intensity (Fig.~3e) and mSHG contrast (Fig.~3f). Note that the mismatch between the experimental data and the fit curves occurs largely around the fundamental SPP resonance, where the resonant $\chi^{(2)}$ contribution used in our simulations is hardly playing any role. 

Based on our numerical calculations, we can conclude that the large value (33$\%$) of
mSHG contrast is dominated by the properties of SPP resonance at
SHG frequency. Thus, upon reversing the external magnetic field the SHG
output can be doubled, which opens up a new strategy for the design
of nonlinear magneto-photonic devices operating at the nanoscale. 
The frequency dependence over the dispersive SPP spectral range clearly demonstrated that the
maximum value of 33$\%$ mSHG contrast remains large
whereas its angular width decreases as the SHG resonance gets
narrower and shifts towards the fundamental resonance.

In the given spectral range one would expect to reach further
enhancement of mSHG contrast by systematically varying the
individual thicknesses in this trilayer structure. The future route for investigations
will imply moving towards the experiments in the telecom frequency range,
which apart from the obvious technological importance would allow to extend the
magneto-optical investigations including the magnetization-induced third
harmonic generation. Moreover, expanding into the far-IR or THz spectral
range, where the SPP dispersion is lower, would allow to explore the
phase-matched magneto-plasmonic coupling between the SPPs at the
fundamental, second, third and higher-order harmonic frequencies.  With the largest reported plasmon-assisted value
of mSHG contrast of 33$\%$ we get closer to the dream of the
nonlinear magneto-photonic devices based on novel physical
principles.


\textbf{Acknowledgements}\\
The authors are indebted to Dr. T.V. Murzina for stimulating discussions and to
the Region Pays de La Loire for funding.
\\

\textbf{Author contributions}\\ All authors have contributed to
this paper and agree to its contents. I.R., V.V.T., and Th.R. wrote the proposal. I.R. and V.V.T. conceived and designed the experiments, performed the measurements and analyzed the data. D.M. and O.G.S. prepared and characterized the samples. A.K, Th.R., D.M., and O.G.S. contributed materials and analysis tools. I.R. and V.V.T. wrote the manuscript. All authors contributed to the discussion of the results.\\

\textbf{Competing interests}\\ The authors declare that they have
no competing financial interests.\\

 \textbf{Correspondence}\\Correspondence and requests for materials should be addressed to
I.R.~(email: razdolski@fhi-berlin.mpg.de) or V.V.T.~(email: vasily.temnov@univ-lemans.fr).





\section*{Supplementary information}

\subsection*{Surface plasmon dispersion in Au/Co/Ag trilayers}

In order to obtain the surface plasmon polariton (SPP) dispersion relation \cite{Raether}
\begin{equation}
\label{SP dispersion} k_{\rm spp}(\omega)=k^{\prime}_{\rm
spp}+ik^{\prime\prime}_{\rm spp}=k_0(n^{\prime}_{\rm
spp}+in^{\prime\prime}_{\rm spp})=
k_0\sqrt{\frac{\varepsilon_{\rm eff}}{1+\varepsilon_{\rm eff}}}\,,
\end{equation}
which is characterized by frequency-dependent SPP wavevector
$k_{\rm spp}(\omega)$ or refractive index $n_{\rm
spp}=n^{\prime}_{\rm spp}+in^{\prime\prime}_{\rm spp}$ we applied
the effective medium theory. It allowed us to approximate the effective
dielectric constant of the Au/Co/Ag interface \cite{Temnov}:
\begin{equation}
\label{e_effective} \varepsilon_{\rm eff}=\frac{1}{\delta_{\rm
skin}}\int_0^{\infty}\varepsilon(z){\rm e}^{-z/\delta_{\rm
skin}}dz\,,
\end{equation}
where the dielectric function $\varepsilon(z)$ depends on the
normal coordinate $z$ inside the multilayer structure. The
spectral dependence of the dielectric function for gold, cobalt
and silver is taken from Ref. \cite{JohnsonChristy}. The total
thickness of two upper metal layers (3~nm Au and 5~nm Co) is
significantly smaller than the penetration depth of the electric
field $\delta_{\rm skin}\sim $12~nm.

In order to explain the results of the graphical representation in
Fig.~1b we would like to quantitatively compare the SPP dispersion
relations at Ag/air and Ag/5nm~Co/3nm~Au/air interfaces. Figure 1S
shows the wavelength dependence of the real ($n^{\prime}_{\rm spp}$) and imaginary ($n^{\prime\prime}_{\rm spp}$) parts of the
SPP refractive index as well as the SPP skin depth and propagation
length $L_{\rm spp}$. The presence of a highly absorbing cobalt layer in our
multilayer structure leads to a significant reduction of the SPP
propagation length $L_{\rm spp}=1/(2k_0n^{\prime\prime}_{\rm
spp})$ over the entire spectral range, as compared to the Ag-air
interface. In contrast to this behavior, the skin depth
$\delta_{\rm skin}$ and the real part of SPP refractive index
$n^{\prime}_{\rm spp}$ exhibit substantial differences only in the
vicinity of the SPP resonance corresponding to an optical
wavelength in vacuum of 350~nm. Moreover, in the entire spectral range of
our SHG measurements (380-460~nm for SHG and 760-920~nm for
fundamental wavelength), the skin depths for both structures
appear to be nearly identical. This allows us to calculate
$\varepsilon_{\rm eff}$ in Eq.~(\ref{e_effective}) using the skin
depth for the Ag-air interface. 

\includegraphics[width=0.9\textwidth]{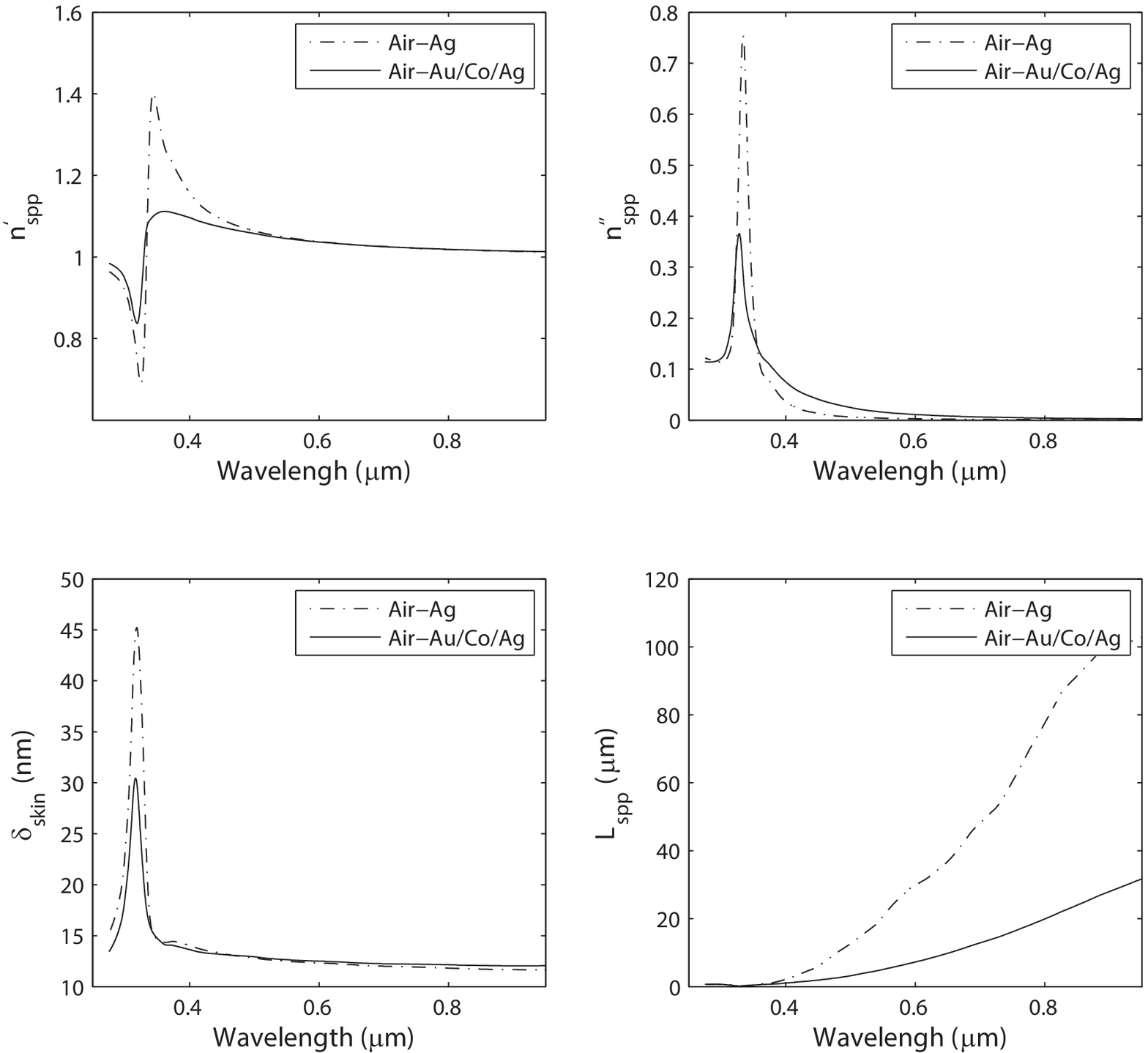}\\
Figure 1S: Spectra dependence of complex refractive index $n_{\rm
spp}=n^{\prime}_{\rm spp}+in^{\prime\prime}_{\rm spp}$, skin depth
$\delta_{\rm skin}$ and propagation distance $L_{\rm spp}$ for
the SPPs excited at the air-Ag and air-Au/Co/Ag interfaces.\\

Now we turn to the discussion of Fig.~1 in the main manuscript.
Due to substantial losses the (linear) resonance condition in
Kretschmann configuration
\begin{equation}
\label{resonance_condition} k_x(\theta,\omega)=k^{\prime}_{\rm
spp}(\omega)\,
\end{equation}
with $k_x(\theta,\omega)=k_0(\omega)n(\omega)\sin\theta$ appears to be
fulfilled within a finite range of incident angles $\theta$ (or
tangential components $k_x(\theta,\omega)$ of optical wavevectors in the
prism).

To account for this angular broadening, we have applied a simple
phenomenological model, where the complex Lorentzian-shaped
function $f(k_x,\omega)$ determines the efficiency of the SPP
excitation:
\begin{equation}
\label{Lorentzian_kx_omega}
f(k_x,\omega)=\frac{k^{\prime\prime}_{\rm
spp}(\omega)}{(k^{\prime}_{\rm
spp}(\omega)-k_x)+ik^{\prime\prime}_{\rm spp}(\omega)}\,,
\end{equation}
or, equivalently
\begin{equation}
\label{Lorentzian_Theta_omega}
f(\theta,\omega)=\frac{k^{\prime\prime}_{\rm
spp}(\omega)}{(k^{\prime}_{\rm
spp}(\omega)-k_0(\omega)n(\omega)\sin\theta)+ik^{\prime\prime}_{\rm
spp}(\omega)}\,.
\end{equation}
The shaded areas in Fig.~1b and Fig.~2c,d in the main manuscript
are limited by two contours of this angular distribution at the
half-maximum of this Lorentzian $|f(k_x,\omega)|^2$ or
$|f(\theta,\omega)|^2$.

\subsection*{Linear reflectivity in presence of SPPs}

Angular dependence of the linear reflectivity in Kretschmann
geometry was measured for eight different wavelengths, namely 760,
800, 860, 920~nm at fundamental laser frequencies and  380, 400,
430 and 460~nm with its second harmonic light (generated in a
BBO-crystal before the sample), see Fig.~2S(a,b). Maximum values
of the reflectivity are normalized to unity because these
supplementary measurements were not calibrated to get absolute
values. Angular spectra of the magneto-optical Kerr effect
measured at the four fundamental frequencies (Fig.~2S,c)
demonstrate great similarity to each other due to low
SPP-dispersion in this spectral region.

Despite the fact that the curves measured in the blue ($2\omega$)
range also look somewhat similar, the change of the resonant dip
when decreasing the wavelength is still noticeable. Figure 2S(c,d)
illustrates angular spectra of the transverse Kerr effect at the
same wavelengths. Note the characteristic asymmetric shape of the
spectra, present for both red and blue sets of wavelengths. It is
worth mentioning that the transverse Kerr effect at blue
wavelengths is several times smaller than at the red ones, which
obviously correlates with the quality factor of the SPP
resonance.
Furthermore, the magnitude of the magnetoplasmonic modulation of the
reflectivity decreases when the wavelength approaches 380~nm. This observation can be
explained by comparing the propagation length of a SPP $L_{\rm
spp}$ with the light wavelength $\lambda$ (see Fig. 1S). It is seen that $L_{\rm
spp}$ decreases together with the fundamental wavelength, thus reducing the quality of
the SPP excitation and also the Co magnetization-induced contribution to the SPP $k$-vector.
As the latter is directly responsible for the SPP-assisted magnetoplasmonic
reflectivity modulation, the transverse magneto-optical Kerr effect also decreases,
albeit a very broad dip in the reflectivity is still seen.

\vspace{0.5cm}
\begin{centering}
\includegraphics[width=0.99\textwidth]{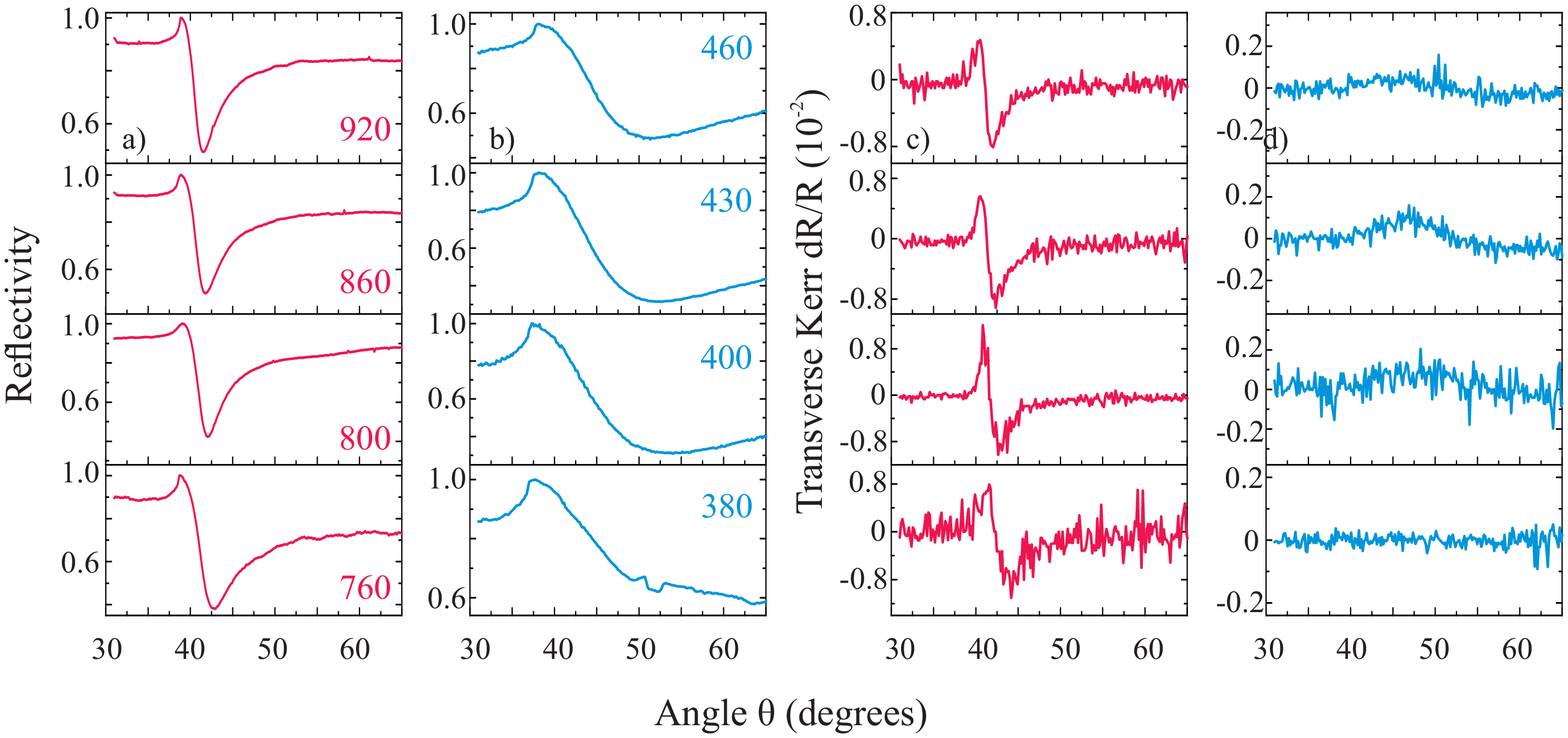}\\
\end{centering}
Figure 2S: (a-b) Angular spectra of the SPP-modulated
reflectivity $R$ for the four red (a) and blue (b) wavelengths.
(c-d) Angular spectra of the magneto-optical transverse Kerr reflectivity modulation $dR/R$ for the same
wavelengths.\\

\subsection*{Nonlinear-optical SPP excitation}

A conventional way of exciting a SPP in the linear optics requires
the linear phase-matching condition (\ref{resonance_condition}) to
be fulfilled \cite{Raether}. This linear phase-matching condition
describes an SPP excitation at a frequency $\omega$. However, this 
phase-matching condition appears to be modified, if an SPP is generated via the nonlinear
frequency mixing, {\it i.e.} surface second harmonic generation (SHG) at the frequency $2\omega$
\cite{Martini,Simon77}. A straightforward analogy with SHG generation in a
nonlinear crystal directly leads to the following phase-matching
condition
\begin{equation}
\label{NLPhaseMatching}
k_x(\omega)+k_x(\omega)=2k_0(\omega)n(\omega)\sin\theta=k^{\prime}_{\rm
spp}(2\omega)\,
\end{equation}
for two incident photons generate generate an SPP at doubled
frequency. As in case of linear resonance this phase-matching
condition is broadened by SPP-dissipation
\begin{equation}
\label{Lorentzian_kx_2omega_nl} f_{\rm
nl}(k_x,2\omega)=\frac{k^{\prime\prime}_{\rm
spp}(2\omega)}{(k^{\prime}_{\rm
spp}(2\omega)-2k_x(\omega))+ik^{\prime\prime}_{\rm
spp}(2\omega)}\,.
\end{equation}
Due to very high plasmonic losses the acceptance angle for
phase-matched SPP generation at $2\omega$ is much larger as
compared to the linear phase-matching at the fundamental frequency.
Note that this nonlinear phase-matching condition slightly differs from the
linear phase-matching condition for a collimated beam at frequency
$2\omega$:
\begin{equation}
\label{Lorentzian_kx_2omega}
f(k_x,2\omega)=\frac{k^{\prime\prime}_{\rm
spp}(2\omega)}{(k^{\prime}_{\rm
spp}(2\omega)-k_x(2\omega))+ik^{\prime\prime}_{\rm
spp}(2\omega)}\,.
\end{equation}
Comparing the expressions for
$2k_x(\omega)=2k_0n(\omega)\sin\theta$ and
$k_x(2\omega)=2k_0n(2\omega)\sin\theta$ we recognize that the
angles for the phase-matched SHG generation and linear excitation in
Kretschmann configuration are different only due to dispersion in
a glass prism: $n(\omega)\neq n(2\omega)$. Knowing this small
angular shift of about 0.7 degree between $f_{\rm
nl}(\theta,2\omega)$ and $f(\theta,2\omega)$ for 860~nm
fundamental wavelength, we can correctly calculate the
SPP-assisted field enhancement at $2\omega$ from the linear
reflectivity in Kretschmann configuration. As we will see below,
this analysis appears to be particularly useful to estimate the
angular dependence of the resonant part of the $\chi^{(2)}$-tensor.

\subsection*{Fitting experimental data}

\subsubsection*{General considerations}

Trying to identify the leading tensor contributions in
second harmonic generation from a isotropic surface is a
delicate issue of playing with three possible contributions 
(consider p-polarised fundamental and SHG beams):
$\chi^{(2)}_{zzz}$, $\chi^{(2)}_{zxx}$ and $\chi^{(2)}_{xzx}$ \cite{Shen}. 
In the chosen Cartesian coordinate frame $z$-axis is perpendicular to
the surface plane and $x$-axis lies in the incidence plane.

There are multiple considerations regarding the appropriate choice
of the nonlinear susceptibility components, if their values are
{\it a priori} not known. The trajectory of free electrons moving
in the direction perpendicular to the interface ($z$-direction) is
expected to exhibit the largest nonlinearities due to the
structural changes across the interface (a large built-in electric
field at the interface explaining the physical origin of the work
function/contact potential) \cite{Raschke}. Therefore the
$\chi^{(2)}_{zzz}$ component is expected to dominate in a
free-carrier approximation, when the energy of SHG photon is
smaller than the energy separation $E_g$ between the Fermi level
and the {\it d}-band. This is indeed the case for silver, as
$E_g=3.8$~eV exceeds the largest photon energy used in our
experiments (3.3~eV). As described in the main Manuscript, there
are two sources of SHG in our model, which are located at the
Ag/glass and Ag/Co/Au/air interfaces. Given that all our
attempts to get reasonable fits with the components other than
$\chi^{(2)}_{1,zzz}$ at Ag/glass interface failed, we fixed this
$zzz$-component in further simulations described below. We
attribute the ability to identify the dominant tensor component to
the aforementioned fact that both relevant frequencies $\omega$
and $2\omega$ are below the d-band edge in silver.

The above conclusions largely rely on the ongoing debates
regarding the choice of the most important components of the
nonlinear susceptibility $\chi^{(2)}$
\cite{Raschke,Petukhov,Sionnest,Naraoka,Heinz,Palomba,Zheng}.
Naraoka {\it et al.} \cite{Naraoka} demonstrated that $\chi_{zzz}$
yields a better agreement with the experimentally observed SHG
output enhancement. The same conclusion is implicitly confirmed by
other authors \cite{Raschke,Heinz}. However, several authors
\cite{Palomba,Zheng} consider instead the $\chi_{xzx}$ component,
following the original paper by Simon {\it et al.} \cite{Simon74},
which nonetheless claims the similar behaviour of all three
components in the vicinity of the SPP resonance. Note the
rigorous, although somewhat less instructive approach used by
Pavlov {\it et al.} \cite{Pavlov}, where all components were taken
into account, and thus the number of the fit parameters increased
dramatically. We attribute these discrepancies to a simple
observation that it is generally difficult to isolate the dominant
contribution to $\chi^{(2)}$ tensor if either the fundamental or
SHG-frequencies overlap with interband transitions.

\includegraphics[width=0.95\textwidth]{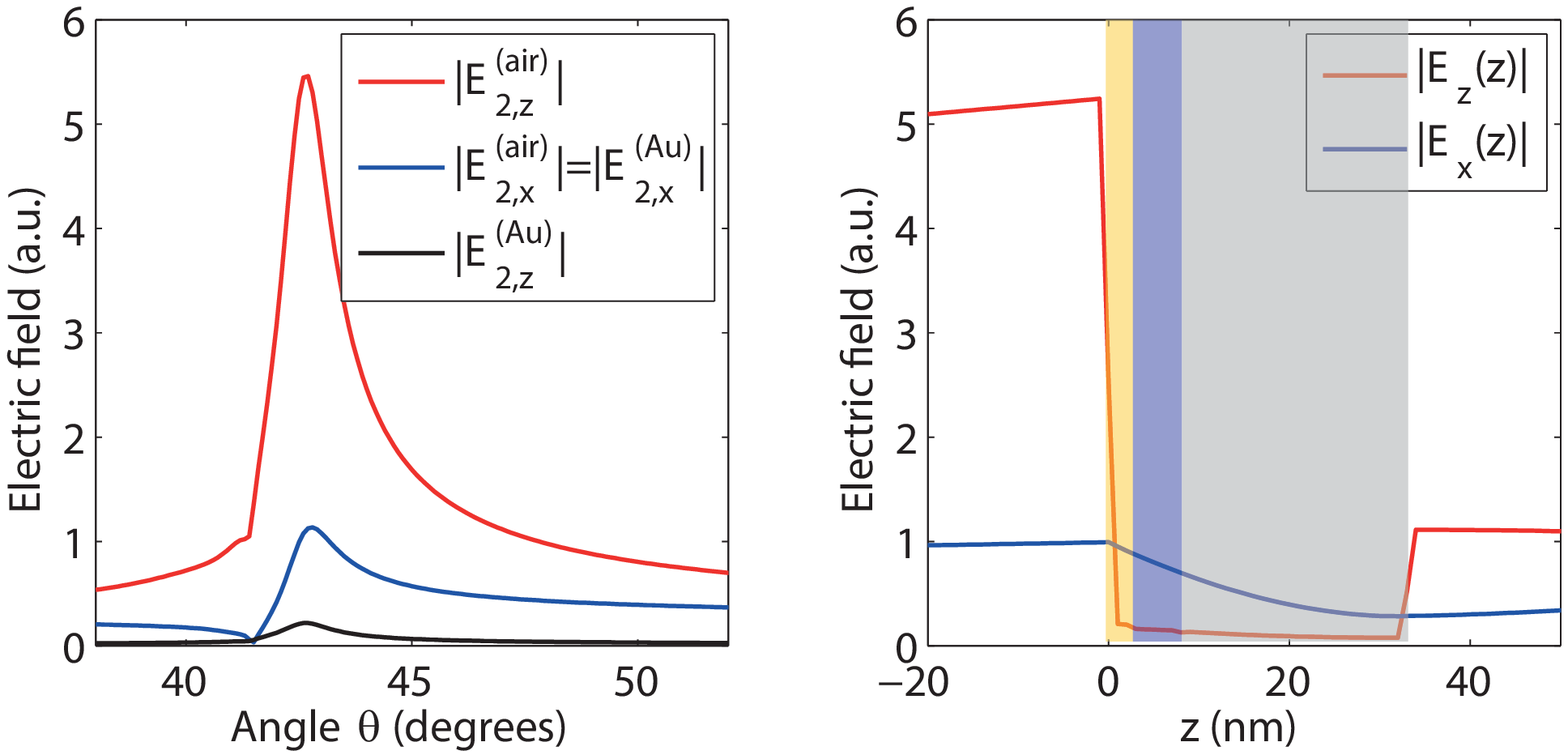}\\
    Figure~3S: (a) Angular dependence of electric fields at the gold-air
    interface. Whereas the tangential component $E_x$ is continuous across the interface,
    the normal component $E_z$ experiences a large jump. (b)
    Dependence of electric field on the normal coordinate $z$ at
    $\theta=42.5^{\circ}$ corresponding to the SPP excitation at the frequency $\omega$.\\

However, different contributions $P^{2\omega}_i=\chi^{(2)}_{ijk}E_jE_k$ to the
non-linear polarization also depend on the components of the local
electric field $E_{j,k}$. The boundary conditions in Maxwell
equations dictate the continuity of the tangential component of
the electric field $E_x$ and the normal component of displacement
field $D_z=\varepsilon(z)E_z$ across the interfaces. In our plasmonic
case characterized by $|\varepsilon_{Au}|\gg 1$ the boundary conditions
suggest $E_z\gg E_x$ in the dielectrics (air and glass) and
$E_z\ll E_x$ in Ag, Au and Co. This statement is confirmed by Finite Difference Time Domain
calculations of the electric fields for our structure leading to
$|E_x|^2\simeq 25 |E_z|^2$ for electric fields {\it inside} the
metal and thus explaining why we have obtained better fits considering the
$\chi^{(2)}_{ixx}E^2_x$ contributions. We would like to note for
the sake of completeness that on the air side the relation is
opposite: $|E_x|^2\simeq 0.04~|E_z|^2$. This suggests that the
nonlinear-optical response of our plasmonic structures covered with a
nonlinear medium (such as, for example, a monolayer of molecules
or semiconductor quantum dots) will be probably dominated by
different contributions to the optical nonlinearity.

\subsubsection*{Fitting SHG intensity}

As a first step in the fitting procedure we decided to try
different $\chi^{(2)}$ components at the top interface, in attempt
to figure out which one is the most important for the observed
modulations of the SHG-signal. In order to make the fitting
procedure more transparent we have limited ourselves to the
following simplified versions of Eq.~(3) in the main Manuscript:
\begin{eqnarray}\label{fit_equations_general}
I^{2\omega}_{zzz}&\propto&|\chi^{(2)}_{1,zzz}E^2_{1,z}\sin\theta+\chi^{(2)}_{2,zzz}E^2_{2,z}\sin\theta|^2\\
I^{2\omega}_{xzx}&\propto&|\chi^{(2)}_{1,zzz}E^2_{1,z}\sin\theta+\chi^{(2)}_{2,xzx}E_{2,x}E_{2,z}\cos\theta|^2\\
I^{2\omega}_{zxx}&\propto&|\chi^{(2)}_{1,zzz}E^2_{1,z}\sin\theta+\chi^{(2)}_{2,zxx}E^2_{2,x}\sin\theta|^2\,,
\end{eqnarray}
where $E_{1},E_{2}$ are the calculated electric fields at the
bottom and top interfaces, respectively, magnetization-induced
$\chi^{(2,m)}$ components are neglected, and $\chi^{(2)}_{2,zzz}$,
$\chi^{(2)}_{2,xzx}$ and $\chi^{(2)}_{2,zxx}$ are the complex
non-magnetic components of the $\chi^{(2)}$-tensor. Additional factors
$\sin\theta$ and $\cos\theta$ originate from the geometrical
projection of the normal ($z$) and tangential ($x$) components of the nonlinear polarisation
on the propagation direction of the SHG-beam in the glass prism (see
Fig.~4S).

\begin{center}\includegraphics[width=0.7\textwidth]{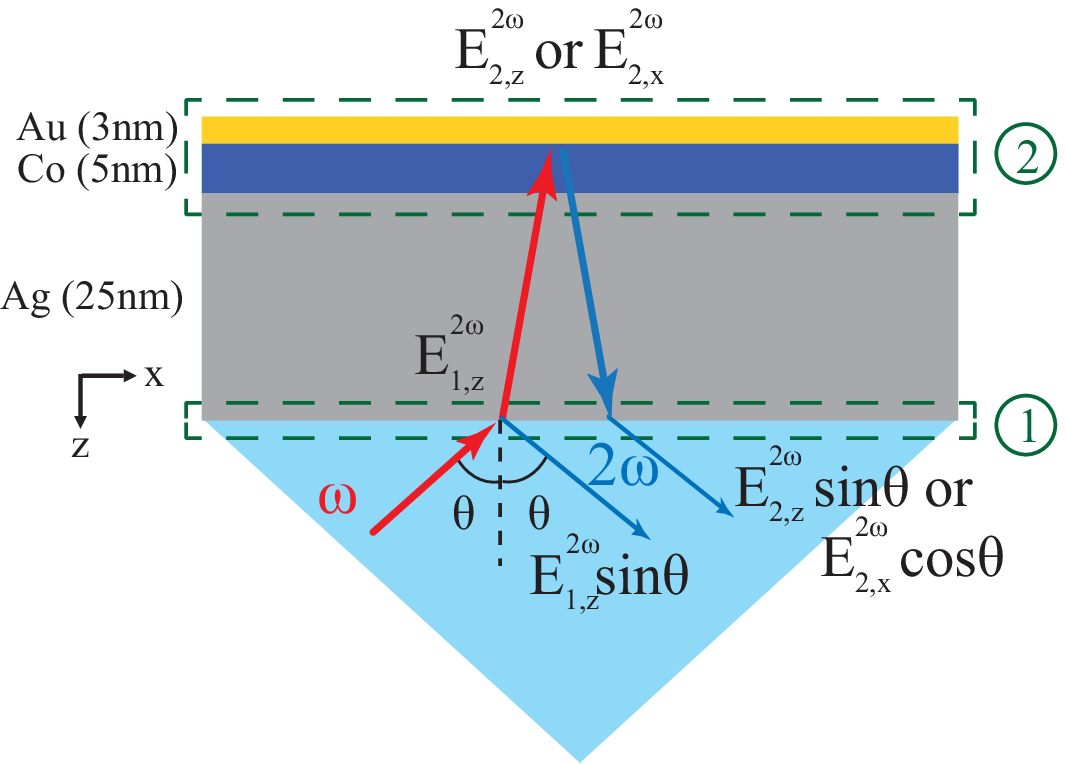}\end{center}
    Figure~4S: Geometrical interpretation of SHG in Au/Co/Ag trilayer structure on a glass prism.
    We assume that SHG occurs at the bottom (1)
    and top (2) interfaces. SHG electric fields in the glass prism depend on SHG
    polarization (normal, linearly polarized along the z-direction or tangential, oscillating along the
    x-direction), which results in a additional geometrical factors
    $\sin\theta$ and $\cos\theta$.\\

These equations can be further simplified in order to minimize the number of the fit parameters:

\begin{eqnarray}\label{fit_equations}
I^{2\omega}_{zzz}&\propto&|E^2_{1,z}\sin\theta+r_{zzz}E^2_{2,z}\sin\theta|^2\\
I^{2\omega}_{xzx}&\propto&|E^2_{1,z}\sin\theta+r_{xzx}E_{2,x}E_{2,z}\cos\theta|^2\\
I^{2\omega}_{zxx}&\propto&|E^2_{1,z}\sin\theta+r_{zxx}E^2_{2,x}\sin\theta|^2\,
\end{eqnarray}

Without the loss of generality we have used complex
coefficients $r_{zzz}$, $r_{xzx}$ and $r_{zxx}$ which represent
the ratios of the respective $\chi^{(2)}_{ijk}$-components to
$\chi^{(2)}_{1,zzz}$ to fit the experimental data. The results of the
best fitting shown in Fig.~4S (a) were obtained using
$r_{zzz}=0.04\cdot{\rm exp}(i1.05\pi)$, $r_{xzx}=0.19\cdot{\rm
exp}(i1.6\pi)$ and $r_{zxx}=0.9\cdot{\rm exp}(i0.05\pi)$, respectively.
Three dashed contours illustrate the Lorentzians given by
expressions (\ref{Lorentzian_kx_omega},\ref{Lorentzian_kx_2omega_nl},\ref{Lorentzian_kx_2omega}). The nonlinear phase-matching
condition shifts the position of SPP-resonance for SHG-generation
to larger angles as compared to the linear SPP resonance at
$2\omega$.

  \includegraphics[width=0.9\textwidth]{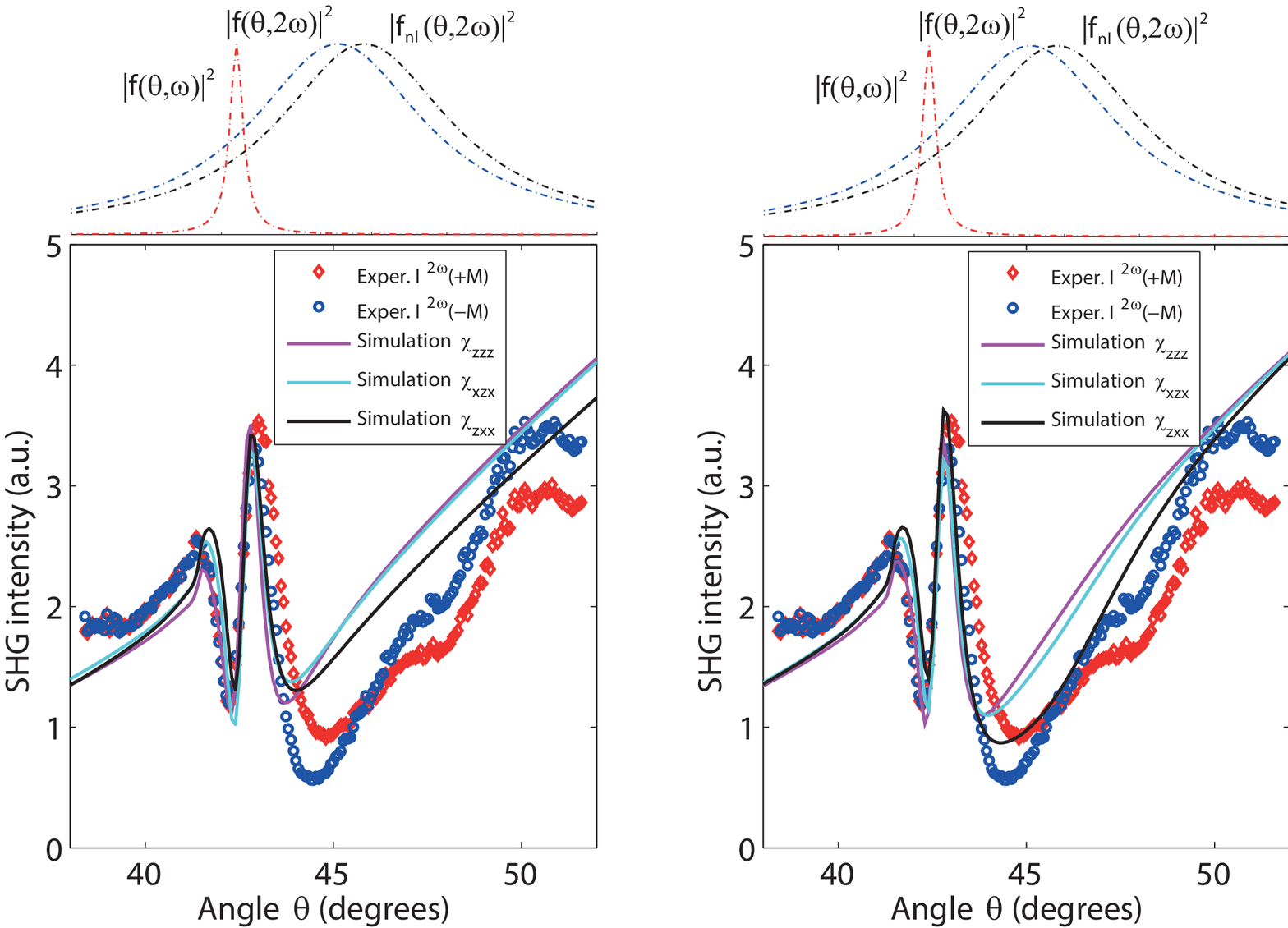}\\
    Figure~5S: Angular dependence of SHG intensity for 860~nm optical pump together
    with best fitting results based on nonresonant $\chi^{(2)}_{\rm nr}$ model (a, see Eq.~(\ref{fit_equations}-14)
    and resonant  $\chi^{(2)}_{\rm res}(\theta)$ model (b, see
    Eq.~(\ref{fit_equations_resonant}-19)).
    Three dashed Lorenzians at the top of the figure aim to illustrate the angular
    position and width of the SPP-resonances at fundamental and SHG frequencies.\\

Two main features can be recognized in Figure 5S(a). First, in the
vicinity of the fundamental resonance ($\theta\simeq~ 42.5^{\circ}$)
all three expressions adequately reproduce the experimental data
and give almost identical fits. Therefore, our fitting attempts
support previously mentioned difficulties and ambiguities
\cite{Simon74} in the identification of the dominant $\chi^{(2)}$ components at
arbitrary metal surfaces.

Second, the behavior of our fit curves around $2\omega$ SPP
resonance ($\theta\simeq~45-46^{\circ}$) is notably different from
the experiment. This failure to reproduce the experimentally
observed SHG intensity minimum stems from the fact that in the
above models we have assumed that $\chi^{(2)}$-tensor components
do not depend on the angle $\theta$. This assumption appears to be
fundamentally wrong when $\chi^{(2)}$ contains resonant (SPP-induced)
contributions at the SHG frequency $2\omega$, thus highlighting one of the
most important features in our modeling of the SPP-assisted SHG.

The usual way to take into account SPP-resonance at SHG-frequency
$2\omega$ is based on the representation of $\chi^{(2)}$-tensor
components as a sum of the non-resonant and resonant contributions
\cite{Shen,Raschke}:
\begin{equation}\label{resonantchi}
\chi^{(2)}=\chi^{(2)}_{\rm nr}+\chi^{(2)}_{\rm res}(\theta).
\end{equation}
In our case the resonant contribution originates from
SPP-excitation at SHG-frequency $2\omega$. Whereas the attempts to
fit the experimental data shown in Fig.~5S(a) were based on the
non-resonant contribution $\chi^{(2)}\simeq\chi^{(2)}_{\rm nr}$
(which does not depend on $\theta$), here we focus on the resonant
part $\chi^{(2)}_{\rm res}(\theta)$. We assume that the angular dependence of the resonant
contribution is as well governed by a Lorentzian shape
\label{resonant_chi2}, which can be obtained from
Eq.~(\ref{Lorentzian_Theta_omega}) by linearizing $\sin\theta$ in
the vicinity of $\theta_{\rm nl}$:
\begin{equation}
\chi^{(2)}_{{\rm
res},ijk}(\theta)=\frac{\Gamma}{\theta-\theta_{\rm nl}+i\Gamma}
r^{\rm (res)}_{ijk}\chi^{(2)}_{1,zzz}\\,.
\end{equation}
Here 
$\Gamma=k^{\prime\prime}_{\rm
spp}(2\omega)/(k_0(\omega)n(\omega)\cos\theta_{\rm
nl})=2.8^{\circ}$, and $\theta_{\rm nl}=45.8^{\circ}$ is given by 
Eq.~(6). 
The expressions for SHG intensity with resonant contributions $\chi^{(2)}_{\rm res}$ thus read:
\begin{eqnarray}\label{fit_equations_resonant}
I^{2\omega}_{zzz}&\propto&|E^2_{1,z}\sin\theta+r^{\rm (res)}_{zzz}f_{\rm nl}(\theta)E^2_{2,z}\sin\theta|^2\\
I^{2\omega}_{xzx}&\propto&|E^2_{1,z}\sin\theta+r^{\rm (res)}_{xzx}f_{\rm nl}(\theta)E_{2,x}E_{2,z}\cos\theta|^2\\
I^{2\omega}_{zxx}&\propto&|E^2_{1,z}\sin\theta+r^{\rm
(res)}_{zxx}f_{\rm nl}(\theta)E^2_{2,x}\sin\theta|^2\,.
\end{eqnarray}
The results of the best fitting in Fig.~5S(b) indicate that all three
equations with $r^{\rm (res)}_{zzz}=0.06\cdot{\rm exp}(i0.3\pi)$,
$r^{\rm (res)}_{xzx}=0.25\cdot{\rm exp}(i0.8\pi)$ and $r^{\rm
(res)}_{zxx}=1.35\cdot{\rm exp}(i1.25\pi)$, respectively, provide a better
approximation to the experimental data. However, the
$I^{2\omega}_{zxx}$ demonstrates the closest proximity to
experimental data thus corroborating our qualitative conclusion that the
contributions containing larger local electric fields
$|E_x|\gg|E_z|$ should dominate.

\subsubsection*{Fitting mSHG and magnetic contrast}

 \includegraphics[width=1.0\textwidth]{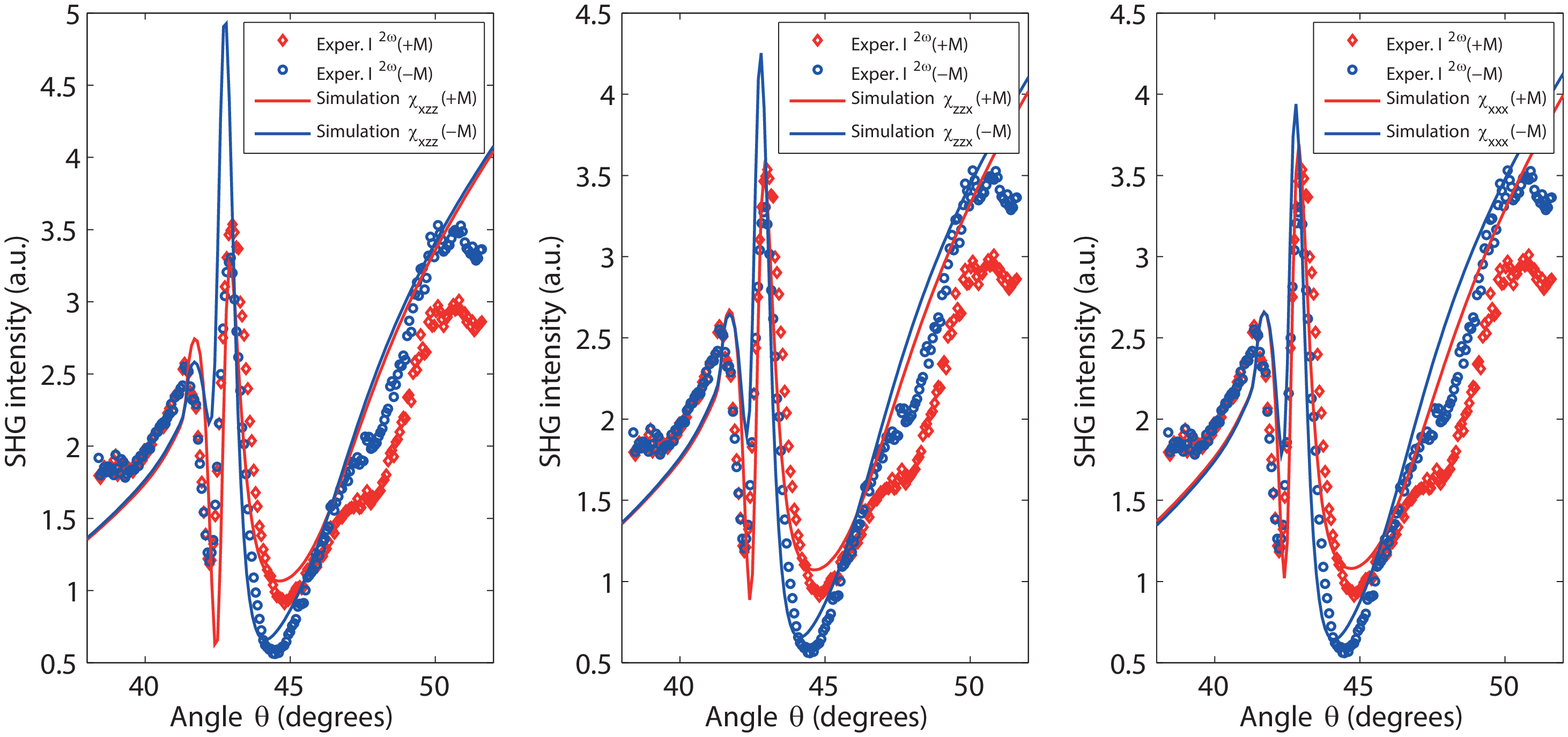}\\
    Figure~6S: Angular dependence of SHG intensity for 860~nm optical pump together
    with best fitting results based on resonant $\chi^{(2)}_{\rm res}(\theta)$ model (see
    Eq.~(\ref{fit_equations_magnetic}-22)) for two opposite orientations of the magnetization $M$.\\

With the results of the two previous sections at hand, we are now
ready to fit our experimental SHG intensities and mSHG contrast to
the Eq.~(3) in the main Manuscript. We note that in the final expression
for the SHG intensity we already know the field distributions
$E_{x,z}$, which account for the SPP resonance at the
fundamental frequency $\omega$, as well as the resonant contribution of
the nonlinear susceptibility $\chi^{(2)}_{\rm res}(\theta)$
associated with the SPP resonance at the double frequency
$2\omega$.

In order to stay in line with systematic fitting procedures we have
considered all three magnetic nonlinear susceptibility components $\chi^{(2,m)}_{xzz}$,
$\chi^{(2,m)}_{zzx}$ and $\chi^{(2,m)}_{xxx}$ \cite{RuPinPan} leading to the
following expressions for the SHG intensities $I(\pm M)$:

\begin{eqnarray}\label{fit_equations_magnetic}
I^{2\omega(m)}_{xzz}&\propto&|E^2_{1,z}\sin\theta+r^{\rm (res)}_{zxx}f_{\rm nl}(\theta)[E^2_{2,x}\sin\theta\pm r^{(m)}_{xzz}E_{2,z}^2\cos\theta] |^2\\
I^{2\omega(m)}_{zzx}&\propto&|E^2_{1,z}\sin\theta+r^{\rm (res)}_{zxx}f_{\rm nl}(\theta)[E^2_{2,x}\sin\theta\pm r^{(m)}_{zzx}E_{2,z}E_{2,x}\sin\theta]|^2\\
I^{2\omega(m)}_{xxx}&\propto&|E^2_{1,z}\sin\theta+r^{\rm
(res)}_{zxx}f_{\rm nl}(\theta)[E^2_{2,x}\sin\theta\pm
r^{(m)}_{xxx}E_{2,x}^2\cos\theta]|^2\,.
\end{eqnarray}

Figure 6S shows the final results of our fitting in a direct
comparison suggesting that  $I^{2\omega(m)}_{xxx}$ provides the
best fitting results, in agreement with our interpretation of
dominating terms proportional to $|E_x(\omega)|^2$. The angular spectra
of the magnetic SHG contrast are presented in Fig.~7S.

\begin{center}\includegraphics[width=0.7\textwidth]{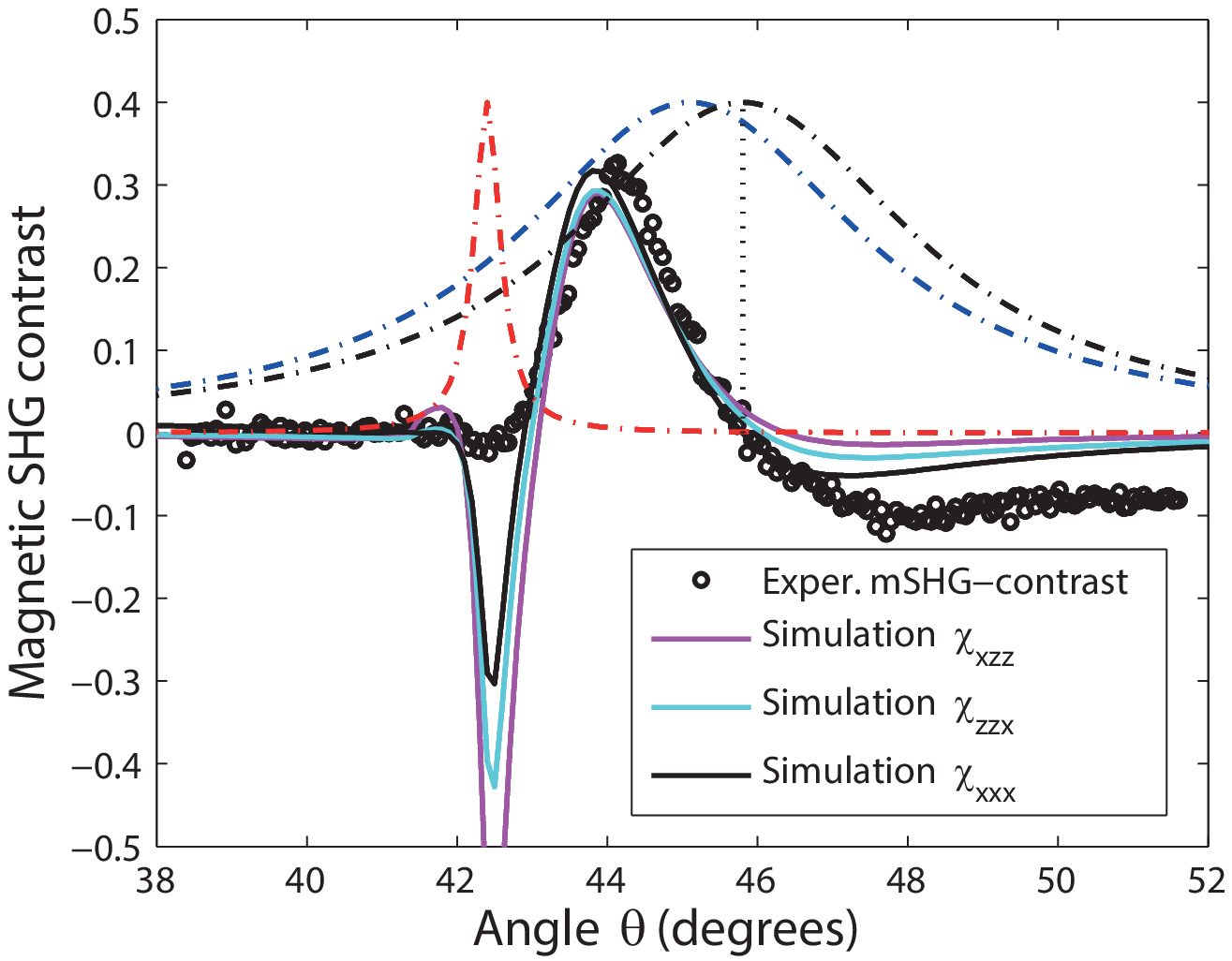}\end{center}
    Figure~7S: Angular dependence of mSHG contrast for 860~nm optical pump together
    with best fitting results based on resonant model with three
    different magnetic components of $\chi^{(2)}_{\rm res}$-tensor(see
    Eq.~(\ref{fit_equations}-9)).
    All fits accurately reproduce the the behavior of the mSHG contrast in the vicinity of the SHG-resonance and,
    notably, the sign change at $\theta_{\rm nl}=45.8$ degrees.\\

The fitting results in Fig.~5S and 6S were obtained with a fixed,
previously deduced value $r^{\rm (res)}_{zxx}=1.35{\rm
exp}(i1.25\pi)$ and using the normalized magnetic components
$r^{(m)}_{ijk}=\chi^{(2,m)}_{ijk}/\chi^{(2)}_{zxx}$ as fit
parameters: $r^{(m)}_{zxx}=0.02\cdot{\rm exp}(-i0.4\pi)$,
$r^{(m)}_{zzx}=0.07\cdot{\rm exp}(i0.05\pi)$ and
$r^{(m)}_{xxx}=0.25\cdot{\rm exp}(i0.5\pi)$. We note that among these
values $r^{(m)}_{xxx}$ is in a good agreement with the results of
theoretical calculations for Fe  predicting a value
$|r_{m}|\simeq0.25$ \cite{Pustogowa}.

The behavior of mSHG contrast in Fig.~7S is accurately
reproduced in the vicinity of the SHG-resonance at $\theta_{\rm
nl}=45.8^{\circ}$ but still shows significant deviations at the fundamental
SPP-resonance around $42.5^{\circ}$. We attribute these
differences to the fact that our simplified model disregards the
non-resonant contributions to the $\chi^{(2)}$-tensor, the inclusion
of which, however, would inevitably stir up troubles related to the
multiparameter fitting. Systematic experimental studies of
similar multilayer structures with different layer thicknesses
should provide more insight about the details of the magnetic SHG
and mSHG contrast behaviour at the fundamental SPP-resonance.

\end{document}